\documentclass[conference]{IEEEtran}
\IEEEoverridecommandlockouts
\usepackage{cite}
\usepackage{amsmath,amssymb,amsfonts}
\usepackage{algorithmic}
\usepackage{graphicx}
\usepackage{textcomp}
\usepackage{xcolor}
\usepackage{makecell}
\usepackage{amsmath,amssymb,amsfonts}
\usepackage{algorithm}  
\usepackage{color}
\usepackage{multirow}
\usepackage{array}
\usepackage{booktabs}
\usepackage{multirow}
\usepackage{fancyhdr} 
\def\BibTeX{{\rm B\kern-.05em{\sc i\kern-.025em b}\kern-.08em
    T\kern-.1667em\lower.7ex\hbox{E}\kern-.125emX}}

\begin{document}

\title{RAN Resource Slicing in 5G Using Multi-Agent Correlated Q-Learning\\}

\author{\IEEEauthorblockN{Hao Zhou, Medhat Elsayed and Melike Erol-Kantarci, \IEEEmembership{Senior Member, IEEE}}
\IEEEauthorblockA{\textit{School of Electrical Engineering and Computer Science} \\
\textit{University of Ottawa}\\
Emails:\{hzhou098, melsa034, melike.erolkantarci\}@uottawa.ca}}
\maketitle
\thispagestyle{fancy} %
      \lhead{} 
      \chead{Accepted by 2021 IEEE International Symposium on Personal, Indoor and Mobile Radio Communications (PIMRC) , \copyright2021 IEEE 
 } 
      \rhead{} 
      \lfoot{} 
      \cfoot{\thepage} 
      \rfoot{} 
      \renewcommand{\headrulewidth}{0pt} 
      \renewcommand{\footrulewidth}{0pt} 
\pagestyle{fancy}

\begin{abstract}
5G is regarded as a revolutionary mobile network, which is expected to satisfy a vast number of novel services, ranging from remote health care to smart cities. However, heterogeneous Quality of Service (QoS) requirements of different services and limited spectrum make the radio resource allocation a challenging problem in 5G. In this paper, we propose a multi-agent reinforcement learning (MARL) method for radio resource slicing in 5G. We model each slice as an intelligent agent that competes for limited radio resources, and the correlated Q-learning is applied for inter-slice resource block (RB) allocation. The proposed correlated Q-learning based inter-slice RB allocation (COQRA) scheme is compared with Nash Q-learning (NQL), Latency-Reliability-Throughput Q-learning (LRTQ) methods, and the priority proportional fairness (PPF) algorithm. Our simulation results show that the proposed COQRA achieves 32.4\% lower latency and 6.3\% higher throughput when compared with LRTQ, and 5.8\% lower latency and 5.9\% higher throughput than NQL. Significantly  higher throughput and lower packet drop rate (PDR) is observed in comparison to PPF.            
\end{abstract}


\section{Introduction}
The forthcoming 5G networks will provide support for vast amount of services and applications, where heterogeneous requirements for latency, bandwidth and reliability will coexist \cite{b1}. Three major traffic types are supported in 5G, namely enhanced Mobile Broad Band (eMBB), Ultra Reliable Low Latency Communications (URLLC), and massive Machine Type Communications (mMTC). The eMBB is regarded as an extension of LTE-Advanced services, which aims to provide high data rate for applications such as video streaming. The URLLC is proposed to provide a sub-millisecond latency and 99.999\% reliability, which is critical for applications such as autonomous vehicles and remote surgery. The mMTC is designed to connect large number of Internet of Things devices, where data transmissions occur sporadically.  

The stringent and heterogeneous QoS requirements of services have become a challenging problem in 5G, especially when different traffic types are multiplexed on the same channel. Considering the limited radio resources and increasing bandwidth demand, different methodologies are proposed for 5G radio resource allocation. A joint link adaptation and resource allocation policy is proposed in \cite{b2}, which dynamically adjusts the block error probability of URLLC small payload transmissions based on cell load. A risk sensitive method is used in \cite{b3} to allocate resources for the incoming URLLC traffic, while minimizing the risk of the eMBB transmissions and ensuring URLLC reliability. Puncturing technique is applied in \cite{b4} to guarantee minimum latency of URLLC, where eMBB traffic is scheduled at the beginning of slots, while URLLC traffic can puncture at any time with a higher priority. 

A common feature of aforementioned works is that URLLC traffic is scheduled on top of eMBB traffic such as puncturing technique\cite{b3,b4}, and a potential priority is applied to guarantee the latency and reliability of URLLC traffic \cite{b2,b3,b4,b5}. As a result, the eMBB traffic will be affected with degraded throughput\cite{b2,b4,b5}. Meanwhile, another important problem is the increasing complexity of wireless networks, e.g., the evolving network architecture, dynamic traffic patterns and increasing devices numbers, which makes it harder to build a dedicated optimization model for resource allocation. 

To this end, the emerging reinforcement learning (RL) techniques become a promising solution\cite{b6}. In \cite{b7}, a Latency-Reliability-Throughput Q-learning algorithm is proposed for jointly optimizing the performance of both URLLC and eMBB users. \cite{b8} develops an RL method to select different scheduling rules according to the scheduler states, which aims to minimize the traffic delay and Packet Drop Rate (PDR). The random forest algorithm is applied in \cite{b9} to accomplish the Transmission Time Interval (TTI) selection for each service, and the result shows a lower delay and lower PDR for URLLC traffic while guaranteeing the eMBB throughput requirements. Furthermore, \cite{b10,b11} use deep reinforcement learning (DRL) scheme for resource allocation in 5G, in which neural networks are used to learn allocation rules.  

In this paper, we propose a multi-agent reinforcement learning (MARL) based resource allocation algorithm, where the performance of URLLC and eMBB are jointly optimized. Different than aforementioned works, we apply the network slicing scheme to aggregate users with similar QoS requirements. Network slicing is an important feature in 5G \cite{b13}. Based on software defined network (SDN) and network function virtualization (NFV) techniques, physical network infrastructures are divided into multiple independent logical network slices. Each slice is presumed to support services with specific QoS requirements, and the whole network achieves a much higher flexibility and scalability. In the proposed correlated Q-learning based inter-slice RB allocation (COQRA) scheme, firstly, each slice is assumed to be an intelligent agent to compete for limited RBs, and the model-free correlated Q-learning algorithm is applied for inter-slice resource allocation. Then resources (more specifically, RBs of the 5G New Radio (NR)) are distributed by each slice among its attached users by proportional fair algorithm, which is the intra-slice allocation \cite{b2}. Compared with Nash Q-learning (NQL) and Latency-Reliability-Throughput Q-learning (LRTQ) techniques \cite{b7}, the results present a 5.8\% and 32.4\% lower latency for URLLC traffic, and 5.9\% and 6.3\% higher throughput for eMBB traffic. COQRA also achieves significantly  higher throughput and a lower PDR than priority proportional fairness (PPF) algorithm. 

The main contribution of this work is that we develop a MARL-based RAN resource slicing scheme for 5G NR. In the proposed multi-agent COQRA, each agent makes decisions autonomously, where they coordinate by exchanging Q-values among each other. Compared with other multi-agent coordination methods, such as Nash equilibrium, the correlated equilibrium is readily solved using linear optimization \cite{b12}, which is critical for the fast response requirement of wireless network.

The rest of this paper is organized as follows. Section \ref{s2} presents related work. Section \ref{s3} defines the system model and problem formulation. Section \ref{s4} introduces the proposed COQRA scheme and the baseline algorithms. Simulation results are presented in section \ref{s5}, and section \ref{s6} concludes the paper.   

\section{Related Work}
\label{s2}
In the literature, various slicing based resource allocation methods have been investigated using both model-free and model-based algorithms. For instance, a QoS framework is proposed in \cite{b14} for network slicing, in which three types of slices are defined. \cite{b15} presents an RL method for resource optimization in 5G network slicing, and the results show an improvement in network utility and scalability. A QoS-aware slicing algorithm is proposed in \cite{b17} where the bandwidth is distributed based on utility function and priority rules. In \cite{b18}, a network slicing and multi-tenancy based dynamic resource allocation scheme is presented, in which hierarchical decomposition is adopted to reduce complexity in optimization. Considering the multi-slice and multi-service scenarios, deep Q-learning is deployed in \cite{b19} for end to end network slicing resource allocation.

Allocation of limited bandwidth resources among slices has been the main challenge in RAN slicing \cite{b20}. The allocation should meet various QoS requirements under the constraint of limited bandwidth budget. Different from existing works, in this paper we solve the resource allocation problem by a MARL approach. We propose a multi-agent COQRA method to distribute RBs among slices. Correlated Q-learning has been applied in microgrid energy management in \cite{b21} to maximize the profit of agents. However, 5G network has much more stringent requirements for agents such as latency and PDR, which is different than the microgrid system. 

\section{System model and problem formulation}
\label{s3}
\begin{figure}[t]
\centering
\includegraphics[width=8.5cm,height=4cm]{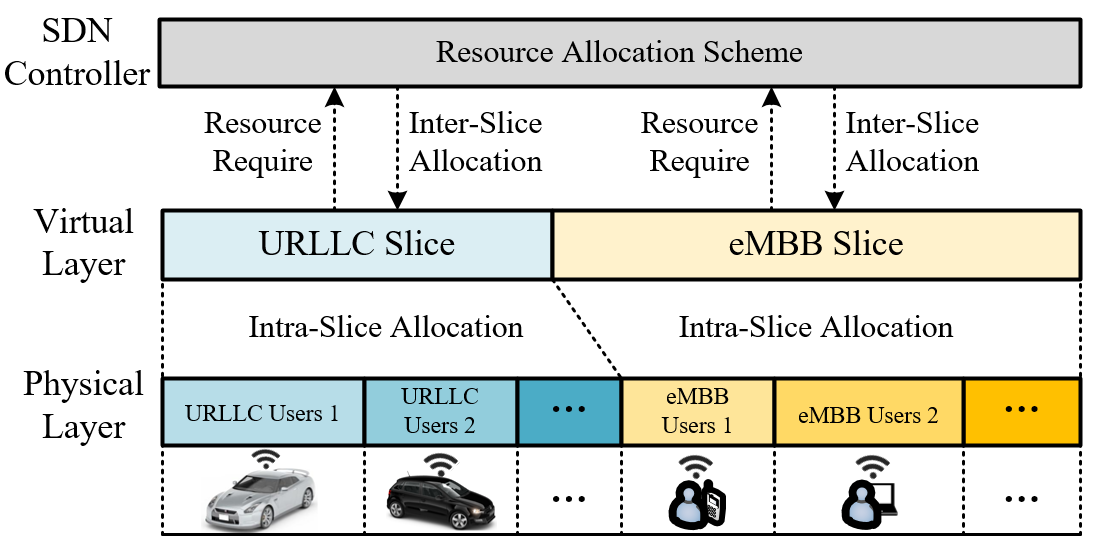}
\caption{Network slicing based two-step resource allocation.}
\label{fig1}
\end{figure}

As shown in Fig.\ref{fig1}, we consider URLLC and eMBB slices where each slice serves several users. First, the slice manager collects QoS requirements of the users such as bandwidth and latency, then the collected information is sent to the SDN controller for the required resources. Based on the received requirements, SDN controller implements the inter-slice RB allocation to distribute RBs between slices. Then, the users are scheduled within the allocated RBs for that particular slice. We consider numerology 0 where one RB contains 12 sub-carriers in frequency domain. 

Here we assume each slice manager is an intelligent agent making decisions autonomously. For the eMBB agent, it needs to maximize the throughput, as denoted by: 
\begin{equation} \label{eu_eqn}
max \sum _{j=1}^J \sum _{e=1} ^{E_{j}}b_{j,e,t},
\end{equation}
where $b_{j,e,t}$ is the throughout of $e^{th}$ eMBB user in $j^{th}$ Base station (BS) at time $t$, and $E_{j}$ is the number of eMBB users of $j^{th}$ BS.

URLLC agent needs to minimize the delay as follows:
\begin{equation} \label{eu_eqn}
min \sum _{j=1}^J \sum _{u=1} ^{U_{j}}d_{j,u,t},
\end{equation}
where $d_{j,u,t}$ is the delay of $u^{th}$ URLLC user in $j^{th}$ BS at time $t$, and $U_{j}$ is the number of URLLC users of $j^{th}$ BS.

The packet delay $d$ mainly consists of three components:
\begin{equation} \label{eu_eqn}
d=d^{tx}+d^{que}+d^{rtx},
\end{equation}
where $d^{tx}$ is the transmission delay, $d^{que}$ is the queuing delay, and $d^{rtx}$ is the HARQ re-transmission delay.

The transmission delay of a packet depends on the link capacity between the UE and the BS:
\begin{equation} \label{eu_eqn}
d^{tx}=\frac{L_{u}}{C_{u,j}},
\end{equation}
where $L_{u}$ is the packet size of $u^{th}$ UE, and $C_{u,j}$ is the link capacity between $u^{th}$ UE and the BS it belongs to. 

The link capacity is calculated as follows:
\begin{small}
\begin{equation} \label{eu_eqn}
C_{u,j}=\sum _{r\in{N^{RB}_{u}}}c_{r}log(1+\frac{p_{j,r}x_{j,r,u}g_{j,r,u}}{c_{r}N_{0}+\sum_{j'\in J_{-j}}{p_{j',r'}x_{j',r',u'}g_{j',r',u'}}}),
\end{equation}
\end{small}
where $N^{RB}_{u}$ is the set of RBs that the $u^{th}$ UE uses, $c_{r}$ is the bandwidth of $r^{th}$ RB, $p_{j,r}$ is the transmission power of $r^{th}$ RB in $j^{th}$ BS, $x_{j,r,u}$ is a binary variable to indicate whether this RB is distributed to $u^{th}$ UE, $g_{j,r,u}$ is the channel gain between BS and UE, $N_{0}$ is the unit noise power density, and $j'\in J_{-j}$ is the BS set except $j^{th}$ BS. 

\section{Correlated Q-learning Based Resource Allocation (COQRA)}
\label{s4}
\subsection{COQRA architecture}
The architecture of the proposed multi-agent COQRA is illustrated in Fig.2. Each slice is an independent agent, and it observes its own state from the environment. The agent exchanges Q-values with its peers to make decisions, and the action selection is determined by correlated equilibrium. Then, the selected actions are sent to wireless environment, and eMBB and URLLC users are scheduled RBs within the slide resources according to the proportional fairness algorithm\cite{b2}. Users transmit packets based on allocated bandwidth, and the experienced throughput and delay are sent back to agents. Finally, the reward is received, and slice managers make next decisions based on new state and updated Q-values.       

\subsection{Markov decision process and Q-learning}
In this section, based on the system model in Section \ref{s3}, we will define the Markov decision process (MDP) to describe agents and introduce the learning scheme. Here we define each slice manager as a intelligent agent, which will interact with the environment and make decisions autonomously.

\begin{figure}[t]
\centering
\includegraphics[width=8cm,height=8cm]{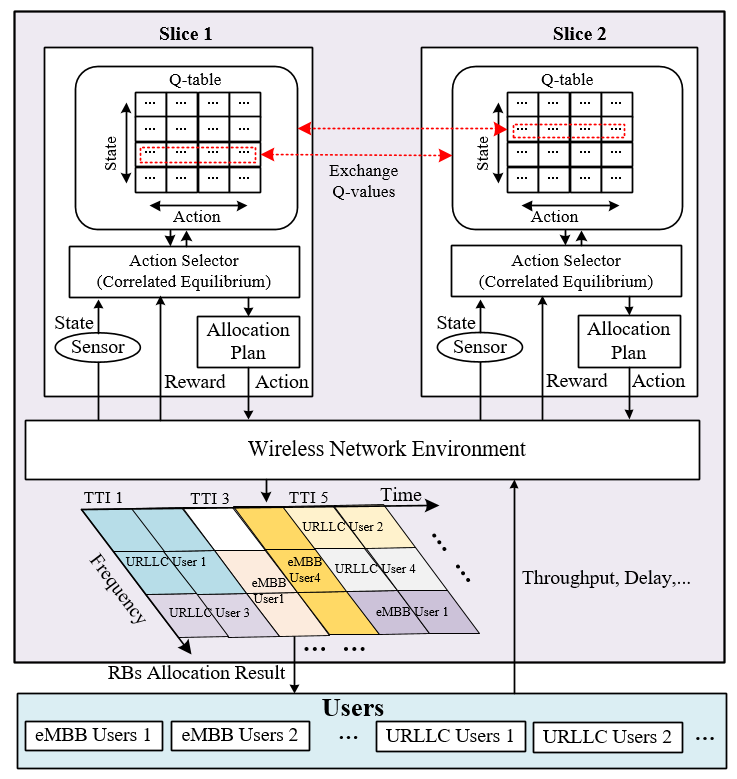}
\caption{Proposed COQRA architecture for intelligent resource management among slices.}
\label{fig}
\end{figure}

We assume each agent has its own state, action and reward signal. The state $s^u$ for URLLC slice manager agent (from hereon, referred to as URLLC agent) is the number of packets in its queue, and the action $a^u$ is the number of RBs it allocates. The state and action for the eMBB slice manager agent (from hereon, referred to as eMBB agent) is defined similarly, namely $s^e$ and $a^e$. Thus, the Q-space for both agents are $Q^u=\{s^u,a^u\}$ and $Q^u=\{s^e,a^e\}$.

The reward function for eMBB agent is given in (\ref{eq7}), where obtaining higher throughput leads to higher reward.
\begin{equation} \label{eq7}
r_{j,t}^{eMBB}=\frac{2}{\pi} \arctan(\sum _{j=1}^J \sum _{e=1} ^{E_{j}}b_{j,e,t}).
\end{equation}

The reward function for URLLC agent at time $t$ is:
\begin{equation} \label{eq6}
r_{j,t}^{URLLC} =
\begin{cases}
  1-\max\limits_{u\in H^{u}_{t}}(d_{u}^{que})^{2}, &   |H^{u}_{t}|\neq 0,\\
0,  &  |H^{u}_{t}|=0,\\
\end{cases} 
\end{equation}
where $d_{u}^{que}$ is the queuing delay for $u^{th}$ URLLC user, $|H^{u}_{t}|$ denotes the length of the queue for URLLC users at time slot $t$. In (\ref{eq6}), a lower queuing delay means a higher reward, which depends on the number of RBs that the agent gets. URLLC agent competes for more RBs to reduce the queuing delay. Meanwhile, to guarantee the PDR performance, we apply a penalty if any packet is dropped.

In Q-learning, one agent always aims to maximize the long-term accumulated reward. For one agent $i$, the state value is:
\begin{equation} \label{eu_eqn}
V^{\pi}_{i}(s_{i}) =\mathbb{E}_{\pi}(\sum_{n=0}^{\infty}\theta^{n} r_{i}(s_{i,n},a_{i,n})|s_{i}=s_{i,0}),
\end{equation}
where $\pi$ is the policy, $s_{i,0}$ is the initial state, $r_{i}(s_{i,n},a_{i,n})$ is the reward of taking action $a_{i,n}$ at state $s_{i,n}$, $\theta$ is the reward discount factor. $V^{\pi}_{i}(s_{i})$ represents long-term expected accumulated reward at state $s_{i}$.

Then we define the state-action value $Q^{\pi}_{i}(s_{i},a_{i})$ to describe the expected reward of taking action $a_{i}$ under state $s_{i}$:
\begin{small}
\begin{equation} \label{15}
Q^{\pi}_{i}(s_{i},a_{i}) = (1-\alpha)Q^{\pi}_{i}(s_{i},a_{i})+\alpha(r(s_{i},a_{i})+\gamma \max Q^{\pi}_{i}(s'_{i},a'_{i}))
\end{equation}
\end{small}
where $s_{i}$ and $s'_{i}$ are current and next state, and $a_{i}$ and $a'_{i}$ are current and next action, and $\alpha$ is the learning rate. By updating the Q-values, the agent will learn the best action sequence, and achieve a long term best reward. 

When there is only one agent, the $\epsilon$-greedy policy is generally applied to balance the exploration and exploitation.
\begin{equation} \label{eu_eqn}
a_{i} =
\begin{cases}
  random \quad action, & rand \leq \epsilon,\\
\arg \max(Q^{\pi}_{i}(s_{i},a_{i})),  &  \text{otherwise},\\
\end{cases}
\end{equation}
where $\epsilon$ is the exploration probability and $0<\epsilon<1$, and $rand$ indicates a random number between 0 and 1. 

On the other hand, in a multi-agent environment, the action of one agent will affect both the environment and the reward of the other agents. Therefore, we propose a correlated Q-learning based resource allocation approach to address the multi-agent 5G environment.

\subsection{Correlated equilibrium}
Given the limited bandwidth resources, in our multi-agent environment, each slice manager agent will compete for more RBs to optimize their own goal, which may lead to a conflict. We use the correlated equilibrium to balance the reward of each agent, and maintain a good overall performance for the whole multi-agent system. In correlated equilibrium, agents exchange Q-values to share information with each other, and the joint action is chosen according to the following equations:
\begin{equation}\label{11}
\begin{aligned}
\max \sum_{\vec a \in A}Pr(\vec s,\vec a)  Q(\vec s,\vec a ) \quad \quad \quad   \quad \quad\\
sub. to \sum_{\vec a \in A}Pr(\vec s,\vec a)=1 \quad \quad \quad \quad \quad \quad \quad\\
\sum_{\vec a_{-i} \in A_{-i}}Pr(\vec s,\vec a_{i})(Q(s,\vec a_{i})-Q(\vec s,\vec a_{-i},a_{i}))\geq0\\
0 \leq Pr(\vec s,\vec a)\leq 1 \quad \quad \quad \quad \quad \quad \quad\\
\end{aligned}
\end{equation}
where $\vec s$ is the system state of eMBB and URLLC agents $\vec s=(s^{e},s^{u})$, $\vec a=(a^{e},a^{u})$ is the joint action, $Pr(\vec s,\vec a)$ is the probability of choosing action combination $\vec a$ under state $\vec s$, $a_{i}$ is the action of agent $i$, $\vec a_{-i}$ is the action combination except agent $i$, and $A_{-i}$ is the set of  $\vec a_{-i}$. The correlated equilibrium is described as a linear program in (\ref{11}). The objective is to maximize the total expected reward of all agents, and the constraints guarantee a probability distribution of action combination in which each agent chooses an optimal action. 

Based on correlated equilibrium, an improved $\epsilon$-greedy policy is applied for action selection:

\begin{equation} \label{eq12}
\pi_{i}(s) =\left\{
\begin{array}{ccl}
  random \quad action, &   rand\leq\epsilon,\\
equation(\ref{11}),  &  rand>\epsilon.\\
\end{array} \right.
\end{equation}
Exploration is performed whenever $rand\leq\epsilon$, i.e., random action is selected. Otherwise the exploitation is implemented by correlated equilibrium. The COQRA scheme is summarized in Algorithm 1. In COQRA, the two-step resource allocation method aviods the complexity of processing all UE requirement by a central controller, which will reduce the computational complexity of learning algorithm by a smaller action space.

\subsection{Nash equilibrium}
In this section, we introduce the NQL algorithm, which is generally used in multi-agent problems. Compared with correlated equilibrium, the Nash equilibrium is a iterative based coordination method. There could be more than one Nash equilibrium or no equilibrium in some cases. We use NQL as a baseline algorithm. In NQL, agents select actions by:
\begin{equation} \label{13}
U_{i}(\vec a_{-i},a_{i})\geq U_{i}(\vec a), \vec a\in A, \vec a_{-i}\in A_{-i},
\end{equation}
where $\vec a$ is the action combination, $\vec a_{-i}$ is the action combination except agent $i$, $A$ is the set of $\vec a$, and $A_{-i}$ is the set of $\vec a_{-i}$. $U_{i}$ is the utility function for agent $i$, which refers to Q-values in this paper. At Nash equilibrium, agents are less likely to change their actions as this will lead to lower observed utility. Similarly, we apply an improved $\epsilon$-greedy policy to select actions as (\ref{eq12}). 
The NQL scheme is summarized in Algorithm 2. We assume the equilibrium is randomly selected if more than one equilibrium are found.  

\begin{algorithm}[!b]
	\caption{COQRA based Resource Allocation}
	\begin{algorithmic}[1]
		\STATE \textbf{Initialize:} Q-learning and wireless network parameters
		\FOR{$TTI=1$ to $T$}
		\IF{$rand<\epsilon$}
		\STATE Choose $a_{t}^{u}$ and $a_{t}^{e}$ randomly.  
		\ELSE
		\STATE Agents exchange Q-values under their current state. Find correlated equilibrium using eq. (\ref{11}) and choose action $a_{t}^{u}$, $a_{t}^{e}$
		\ENDIF
		\STATE Complete the inter-slice resource allocation.
		\STATE Implement intra-slice allocation. Schedule users on respective slice resources using the proportional fair algorithm.
		\STATE Agents calculate reward based on received QoS metrics. 
		\STATE Update agent state $s^{u}$, $s^{e}$ and Q-table $Q^u$ and $Q^e$ .   
		\ENDFOR
	\end{algorithmic}
\end{algorithm}

\setlength{\textfloatsep}{3pt}
\begin{algorithm}[!b]
	\caption{NQL based Resource Allocation}
	\begin{algorithmic}[1]
		\STATE \textbf{Initialize:} Q-learning and wireless network parameters
		\FOR{$TTI=1$ to $T$}
		\IF{$rand<\epsilon$}
		\STATE Choose $a_{t}^{u}$ and $a_{t}^{e}$ randomly.  
		\ELSE
	        \FOR{Each agent}
		      \STATE Search its own Nash equilibrium using eq. (\ref{13}).
		     \ENDFOR
		      \STATE Match each agent's equilibrium to get the system Nash equilibrium.
		  \STATE Agents choose actions according to system Nash equilibrium.
		\ENDIF
		\STATE Complete the inter-slice resource allocation.
		\STATE Implement intra-slice allocation, using the proportional fair algorithm.
		\STATE Agents calculate reward based on received QoS metrics. 
		\STATE Update agent state $s^{u}$, $s^{e}$ and Q-table $Q^u$ and $Q^e$.   
		\ENDFOR
	\end{algorithmic}
\end{algorithm}

\subsection{LRTQ and PPF algorithms}
To further investigate the performance of the proposed method, two more baseline algorithms are used in this paper. LRTQ was proposed in \cite{b7}. LRTQ is also a learning-based resource allocation method, but it only defines one reward function for all users. PPF algorithm is applied in \cite{b2}. In PPF, the RBs are first allocated to URLLC users to guarantee low latency, then the remaining RBs are distributed among eMBB users. Note that network slicing is not implemented in both of these baseline algorithms. These algorithms perform RB allocation decisions only.

\begin{figure}[!b]
\vspace{-5pt}
\centering
\includegraphics[width=7cm,height=5cm]{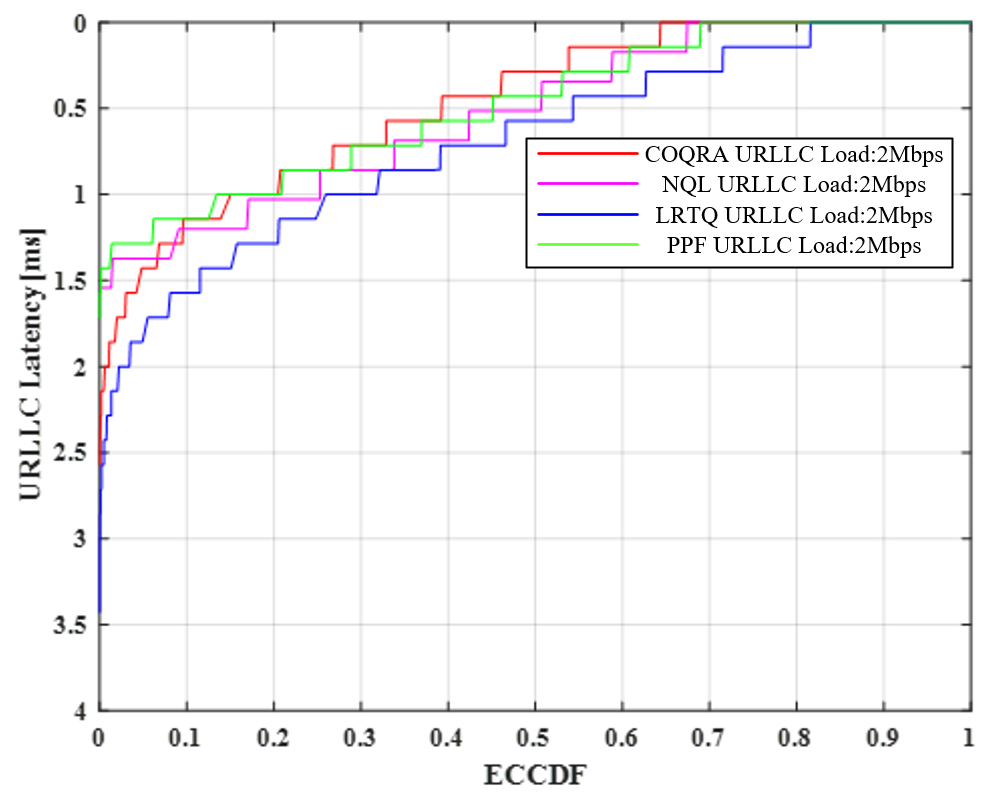}
\setlength{\abovecaptionskip}{-2pt} 
\caption{URLLC latency distribution[ms] under load=2Mbps per cell.}
\label{fig2}
\end{figure}

\begin{figure}[!b]
\centering
\includegraphics[width=7cm,height=5cm]{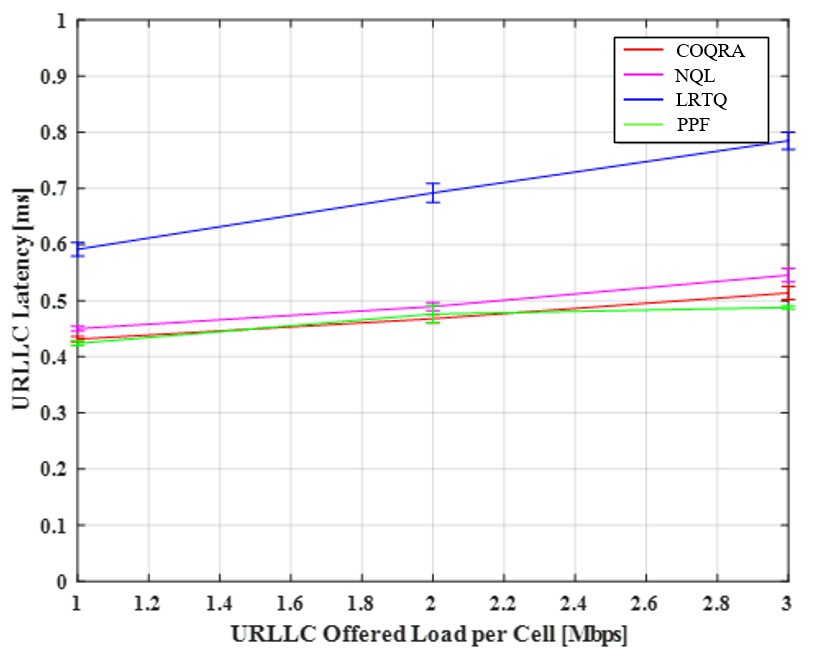}
\setlength{\abovecaptionskip}{-2pt} 
\caption{URLLC latency [ms] with varying offered load [Mbps] per cell.}
\label{fig3}
\end{figure}

\section{Performance evaluation}
\label{s5}
\subsection{Parameter settings}

\begin{table}[t]
\caption{Parameters Settings}
\centering
\renewcommand\arraystretch{1.4}
\begin{tabular}{|p{2.5cm}<{\centering}|p{5cm}<{\centering}|}
\hline
Parameters & Value\\
\hline
\multirow{3}*{Traffic Model} & \makecell[c] {URLLC Traffic: 80\% Poisson distribution\\ and 20\% constant bit rate.} \\
\cline{2-2}
~ & eMBB Traffic: Poisson distribution.  \\
\cline{2-2}
~ & Packet Size: 32 Bytes.\\
\hline
\multirow{2}*{Propagation Model} & 128.1+37.6log(distance(km)). \\
\cline{2-2}
~ & Log-Normal shadowing with 8 dB. \\
\hline

\multirow{3}*{Transmission settings} & \makecell[c] {Transmission power: 40 dBm\\ (Uniform distributed)}\\
\cline{2-2}
~&Tx/Rx antenna gain: 15 dB. \\
\cline{2-2}
~ & 3GPP urban network.  \\
\hline

\multirow{3}*{Q-learning} & Learning rate: 0.9\\
\cline{2-2}
~ &Discount factor: 0.5 \\
\cline{2-2}
~ & Epsilon value: 0.05\\
\hline
\end{tabular}
\label{1}
\vspace{-6pt}
\end{table}

We use MATLAB to implement our proposed algorithm. We consider five gNBs with 500 meter inter-site distance, each serving one eMBB and one URLLC slice. Each eMBB slice is serving 5 users, and URLLC slice has 10 users, which is randomly distributed in the cell. The bandwidth for a cell is 20 MHz, and there are 100 RBs. Each RB contains 12 sub-carriers, and each sub-carrier has 15kHz. 100 RBs are divided into 13 RB groups, where the first 12 groups contain 8 RBs each, and the last group contains 4 RBs. The simulation period is 5000 TTIs, and each TTI contains 2 OFDM symbols (5G mini-slot length 0.143ms)  allocations. The $\epsilon$-greedy policy is implemented in first 3000 TTI, and the rest 2000 TTI is pure exploitation. Other parameters are shown in Table \ref{1}. Each scenario is repeated 10 runs to get an average value with 95\% confidence interval.

\subsection{Simulation Results }

\begin{figure}[!b]
\vspace{-5pt}
\centering
\includegraphics[width=7cm,height=5cm]{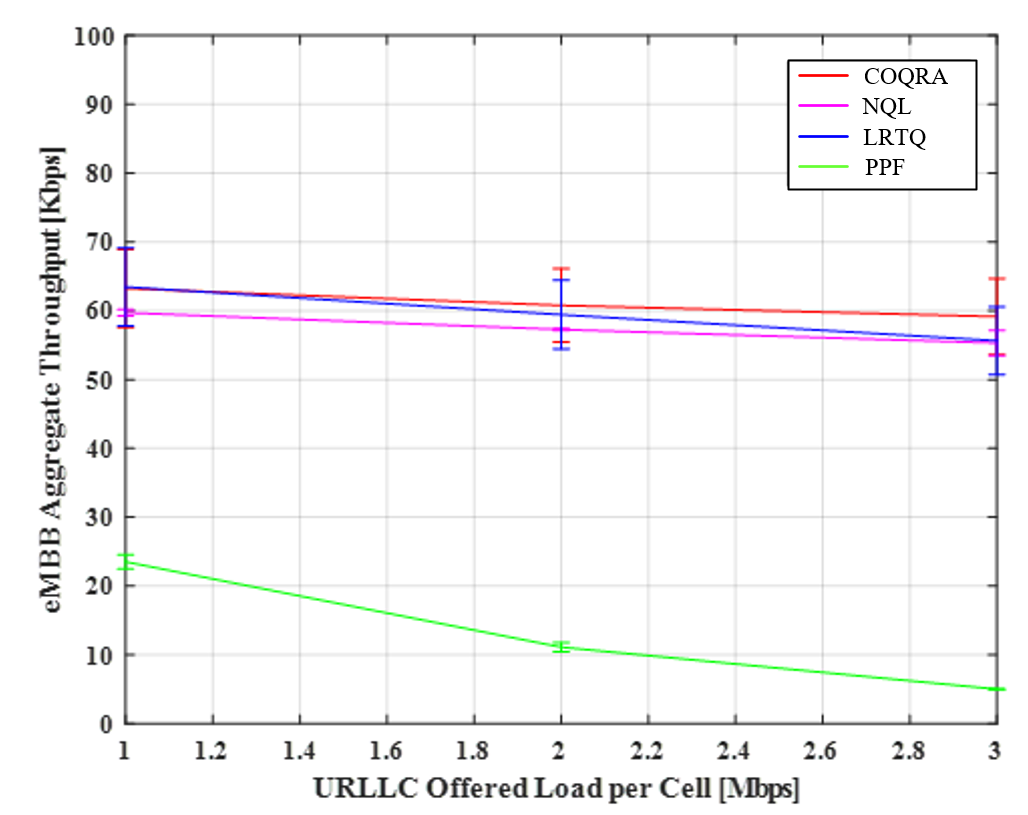}
\setlength{\abovecaptionskip}{-2pt} 
\caption{eMBB throughput with varying URLLC loads [Mbps].}
\label{fig4}
\end{figure}

\begin{figure}[!b]
\centering
\includegraphics[width=7cm,height=5cm]{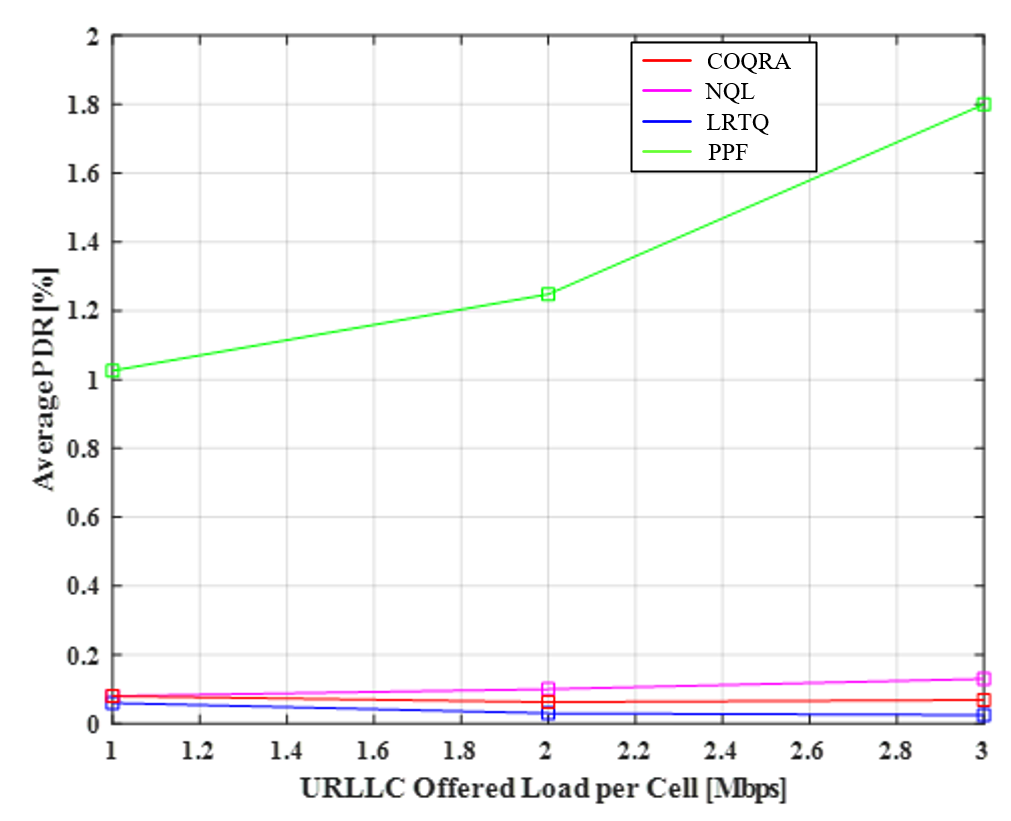}
\setlength{\abovecaptionskip}{-2pt} 
\caption{Average PDR under varying URLLC loads [Mbps].}
\label{fig5}
\end{figure}

First, we set eMBB load to 0.5 Mbps per cell, and consider that URLLC load changes from 1 Mbps to 3 Mbps per cell. The latency distribution of four algorithms are shown in Fig.\ref{fig2}. Meanwhile, Fig.\ref{fig3} presents the average URLLC latency against varying URLLC offered loads. The results show that COQRA, NQL and PPF have a comparable latency distribution, while the LRTQ has a relatively higher latency. The PPF has the lowest overall latency for URLLC traffic, and the main reason is that URLLC traffic has a priority in this method. Whenever URLLC packet arrives, it will be directly scheduled over eMBB traffic in the PPF algorithm. Meanwhile, the COQRA achieves 4.4\% and 5.8\% lower latency than NQL when URLLC load is 2 Mbps and 3 Mbps, respectively. Compared with LRTQ, the COQRA has a 27.1\% lower latency under 2 Mbps load, and 32.4\% lower latency under 3 Mbps load. 

\begin{figure}[!t]
\centering
\includegraphics[width=7cm,height=5cm]{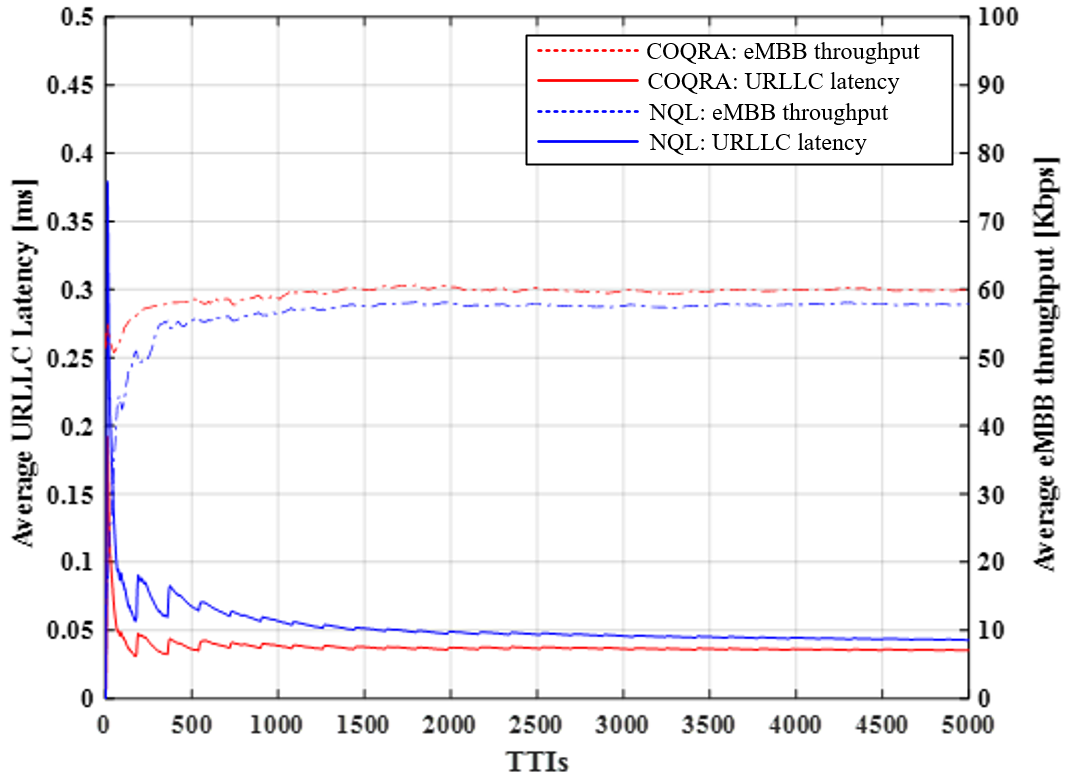}
\setlength{\abovecaptionskip}{-2pt} 
\caption{Convergence performance of COQRA and NQL.}
\label{fig6}
\vspace{-5pt}
\end{figure}

Furthermore, the eMBB throughput under different URLLC load is shown in Fig.\ref{fig4}. The result shows that COQRA, NQL and LRTQ have a close performance of throughput, while the PPF has a much lower value. When URLLC load is 3 Mbps, the COQRA method has a 5.9\% higher throughput over NQL method, and 6.3\% higher than LRTQ. Although the PPF has a good latency performance for URLLC, the eMBB throughput is almost 90\% lower than other three algorithms. This result can still be explained by the priority settings in PPF, which means the eMBB throughput will decrease with increasing prioritized URLLC load. On the other hand, the COQRA, NQL and LRTQ benefit from the jointly optimizing scheme, and they maintain a good throughput performance.  
In Fig.\ref{fig5}, we compare the PDR of four algorithms. We show that COQRA, NQL and LRTQ maintain a much lower PDR than PPF method. In learning algorithms, agents will get a negative reward when dropping packets. However, PPF fails to maintain low PDR under all traffic loads, where a worst case PDR of $1.8\%$ is observed. Finally, we compare the convergence performance of COQRA and NQL, and a faster convergence is observed for COQRA in Fig.\ref{fig6}. The reason is that COQRA has a more efficient way to find the equilibrium. Overall, COQRA outperforms all baseline methods in terms of latency, throughput, packet loss and convergence time.

\section{Conclusion}
\label{s6}
5G and beyond 5G networks will serve heterogeneous users of multiple slices which calls for new ways of network slicing and resource allocation. Machine learning techniques provide a promising alternative to the existing schemes. In this paper, we propose a Radio Access Network (RAN) slicing based resource allocation method for 5G, namely correlated Q-learning based inter-slice RB allocation (COQRA), to allocate radio resources to eMBB and URLLC slices. The proposed algorithm is compared with Nash Q-learning method, Latency-Reliability-Throughput Q-learning method and priority proportional fairness algorithm. Simulation results show that the proposed COQRA scheme achieves the best overall performance. In the future works, we plan to enhance the scalability of COQRA such that it can be used for intra-slice allocations.

\section*{Acknowledgment}
We thank Dr. Shahram Mollahasani for his generous help and useful discussions. This work is funded by the NSERC CREATE and Canada Research Chairs programs.

\end{document}